\documentclass[fleqn,twoside]{article}
\usepackage{espcrc2}

\usepackage{graphicx}
\usepackage[figuresright]{rotating}

\newcommand{\rf}[1]{(\ref{#1})}
\newcommand{\beq}{\begin{equation}}
\newcommand{\eeq}{\end{equation}}
\newcommand{\bea}{\begin{eqnarray}}
\newcommand{\eea}{\end{eqnarray}}

\newcommand{\e}{\mbox{e}}
\renewcommand{\d}{\mbox{d}}

\renewcommand{\L}{\Lambda}
\renewcommand{\b}{\beta}
\renewcommand{\a}{\alpha}

\newcommand{\Del}{\Delta}

\newcommand{\tr}{\mbox{Tr}\;}

\newcommand{\mi}{\!-\!}
\newcommand{\equ}{\!=\!}
\newcommand{\pl}{\!+\!}

\newcommand{\AmS}{{\protect\the\textfont2
  A\kern-.1667em\lower.5ex\hbox{M}\kern-.125emS}}

\title{3d Lorentzian, Dynamically Triangulated Quantum Gravity}

\author{J. Ambj\o rn\address[MCSD]{Niels Bohr Institute,  
        Blegdamsvej 17, DK-2100 Copenhagen \O, Denmark}%
       \thanks{Supported by MaPhySto,  Center for 
Mathematical Physics and Stochastics -- financed by the 
Danish National Research Foundation}{${}^\ddag$},
J. Jurkiewicz\address{Institute of Physics, Jagellonian University,
Reymonta 4, PL 30-059 Krakow, Poland}{${}^*$}\thanks{Supported
by KBN grant 2P03B 01917}{${}^\ddag$} and
R. Loll\address{Institute for Theoretical Physics, Utrecht University, 
Leuvenlaan 4, NL-3584 CE Utrecht, \\
The Netherlands}\thanks{
Supported by EU network on ``Discrete Random Geometry'', 
grant HPRN-CT-1999-00161, 
and by ESF network no.82 on ``Geometry and Disorder''.}
}
       
\begin{document} 

\begin{abstract}
The model of Lorentzian three-dimensional dynamical triangulations provides
a non-perturbative definition of three-dimensional quantum gravity.
The theory has two phases:  a weak-coupling phase with  quantum fluctuations
around a ``semiclassical'' 
background geometry which is generated dynamically despite 
the fact that the formulation is explicitly background-independent,
and a strong-coupling phase where ``classical'' space disintegrates
into a foam of baby universes. 
\vspace{1pc}
\end{abstract}

\maketitle

\section{Introduction}

Three-dimensional quantum gravity provides a good model
for the study of quantum gravity. Locally, the classical solution 
is just flat space, or in the case of a positive cosmological 
constant, 3d deSitter space. If one expands around such 
a classical solution in order to quantize the theory
it is non-renormalizable. Nevertheless we know the theory 
has no dynamical {\it field degrees of freedom}, but only
a finite number number of degrees of freedom. Thus it can 
be quantized following different procedures, e.g.\ reduced 
phase space quantization. However, it remains unclear if 
anything is ``wrong'' in an approach where one perform the 
summation over fluctuating three-geometries and how 
such an approach deals with the seemingly non-renormalizability
of the theory of fluctuating geometries. 

In two-dimensional quantum gravity the simplest direct 
approach to the theory of fluctuating geometries, known 
as {\it dynamical triangulations} has 
been very successful (see e.g.\\ \cite{jan2d}).
However, simple generalizations to higher dimensions seem
not to work. In four-dimensional space-time the failure 
could in principle be due to the non-existence of a four-dimensional 
theory of quantum gravity which is not embedded in a larger 
theory, but this argument does not apply to the three-dimensional
theory of quantum gravity.  This led two of us to suggest,
following an old idea by Teitelboim, that Euclidean quantum 
gravity might not be related to the ``real'' Lorentzian 
quantum gravity and that one should only include 
causal geometries in the sum over histories \cite{al}. 
The geometries which 
appear in the regularized 
version of such a theory were called {\it Lorentzian dynamical 
triangulations}, and each of these geometries has a well defined 
rotation to an Euclidean geometry. The opposite is not true 
however: there are many Euclidean geometries which cannot be 
rotated to a Lorentzian geometry with a global causal structure.
However, it implies that one {\it can} perform the summation
over this restricted class of geometries in the ``Euclidean sector'',
and rotate back after the summation has been done. This is 
the way we will treat the summation over histories in the following. 
 
In two dimensions one can perform the summation over the 
class of Lorentzian geometries explicitly and  obtain 
a theory which is {\it different} from Euclidean
two-dimensional quantum gravity. The difference is 
best illustrated by considering what is called the proper-time 
propagator, where one sums over all geometries with 
the topology $S^1\times[0,1]$, where 
the two spatial boundaries are 
separated by a proper time $T$. 
In the Lorentzian theory the spatial  slices at a time $T' < T$ 
are characterized by the fact that the spatial topology 
always remain a circle. In two-dimensional Euclidean quantum gravity 
similar spatial slices corresponding to constant 
proper time  split up into  many (in the continuum limit
infinitely many) disconnected ``baby'' universes, {\it each} having 
the topology $S^1$.

\section{3d Lorentzian dynamical triangulations}  

\subsection{Theory}   

As mentioned above the proper-time propagator is a 
convenient object to study. We choose the simplest possible 
topology of space-time, $S^2\times [0,1]$, so that the spatial 
slices of constant proper time have the topology of a two-sphere.
Each spatial slice has an induced two-dimensional Euclidean
geometry. In the formalism of Lorentzian dynamical triangulations
the space of Euclidean 2d geometries is approximated by 
the space of 2d dynamical triangulations. 
In order to obtain a three-dimensional triangulation of 
space-time we have to fill space-time between two successive
time slices. This is done as follows: above (and below) each 
triangle at proper time $t\equ n\, a$, $n$ integer, 
we erect a tetrahedron with its 
tip at $t\pl a$, so-called (3,1)-tetrahedra ($t\mi a$, so-called
(1,3)-tetrahedra). Two tetrahedra which share a 
common spatial link in the $t$-plane might be glued together along 
a common time-like triangle. Remaining free time-like triangles
are glued together by so-called (2,2)-tetrahedra which have 
a spatial link both in the $t$-slice and the $t\pl a$ slice.
(2,2)-tetrahedra can also be glued to each other in 
all possible ways, the only restriction being that 
if we cut the triangulation in a constant $t$-plane 
between $t$ and $t\pl a$, the corresponding graph, which 
consists of triangles and squares (coming from cutting the 
(2,2)-simplexes) forms a graph with spherical topology.

Summing over all such piecewise linear geometries with the 
Boltzmann weight given by the Einstein-Hilbert action defines
the sum over geometries (see \cite{ajl} for details).
The partition function becomes (up to boundary terms)
\beq\label{wick2}
Z(k_0,k_3,T)=\sum_{T} 
\e^{k_0 N_0(T)-k_3N_3(T)},
\eeq
where the summation is over the class of triangulations mentioned,
and $N_0(T)$ denotes the total number of vertices and $N_3(T)$
the total number of tetrahedra. $k_0$ is inversely proportional 
to the bare inverse gravitational coupling constant, while 
$k_3$ is linearly related to the cosmological coupling constant.

\subsection{Numerical simulations}

\begin{figure}[htb]
\vspace{-2.5cm}
\includegraphics[width=7.5cm]{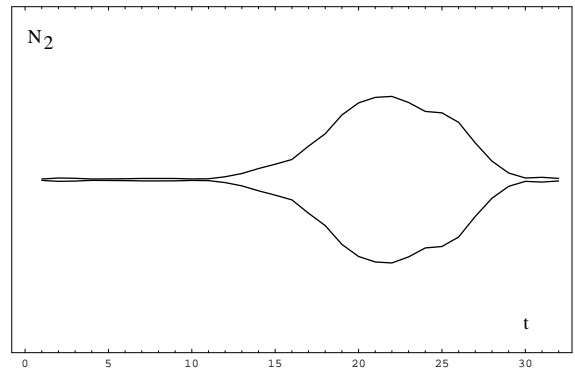}
\vspace{-3.5cm}
\caption[fig1]{a snapshot of a three-dimensional configuration of 
space-time. The vertical axis is the spatial volume $N_2$, the 
horizontal axis the  proper time $t$}
\label{fig1}
\end{figure}

The model \rf{wick2} can be studied by Monte Carlo simulations
(see \cite{ajl} for details). There is only one phase\footnote{In
some previous studies we observed a phase transition for large 
$k_0$. This was caused by restrictions on the gluing of 
(2,2)-simplexes. We have now dropped these restrictions.}.
Let us fix the total three-volume of space-time to be $N_3$,
and let us take the total proper time $T$ large. 
One observes the appearance of a ``semiclassical'' lump of universe,
as shown in Fig.\ \ref{fig1}.
We observe that a typical  spatial volume $N_2(t)$ in 
the lump, and the time-extent $\Del T$ of the lump scale as:
\beq\label{scale}
N_2 \sim N_3^{2/3},~~~~\Del T \sim N_3^{1/3}.
\eeq
This justifies the use of the word ``semiclassical'' for the lump.
In the computer simulations the center of mass of the lump 
moves around randomly and there are  fluctuations in the 
spatial volumes. We have studied the fluctuations of 
{\it successive} spatial volumes and the distributions 
of such spatial volumes is very well described by 
\beq\label{distrib}
P(N_2(t),N_s(t+a)) \sim 
\e^{ -c(k_0) \frac{(N_2(t\pl a)-N_2(t))^2}{N_2(t\pl a)+N_2(t)}}.
\eeq
The constant $c(k_0)$ decreases as $k_0$ increases (i.e.\ the 
bare gravitational coupling constant decreases). At the same 
time one can observe that the total number of (2,2)-simplices
decreases. The (2,2)-simplices act as glue between successive 
spatial volumes.

Thus an effective action for the spatial volume of the model
seems to be
\beq\label{sitter} 
S_{eff}(V_2) = \int dt \; \Big( \frac{\dot{V}_2^2(t)}{V_2(t)} + \L V_2(t)\Big).
\eeq
This is exactly the classical action for the spatial 
volume in proper-time gauge, thus supporting the 
semiclassical interpretation of the lump.

\section{3d Lorentzian gravity as a matrix model}
\begin{figure}[htb]
\vspace{-2cm}
\centerline{\includegraphics[width=8cm,angle=-90]{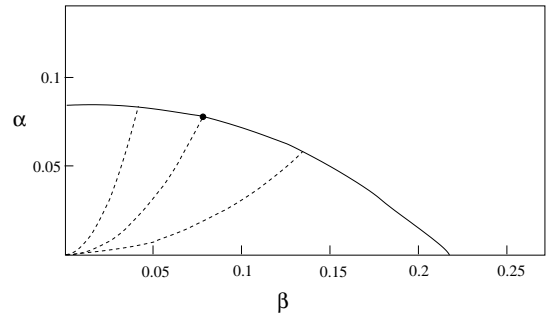}}
\vspace{-2.8cm}
\caption[fig4]{The phase diagram of 3d Lorentzian gravity from matrix models,
calculated in \cite{kz}}
\label{fig4}
\end{figure}

If we slice our three-dimensional configurations, not at proper
time $t$, but at proper time $t\pl a/2$ we will, as mentioned 
earlier, obtain a spherical graph with from two types of triangles,
coming the spatial intersections of (1,3)- and (3,1)-tetrahedra, respectively.
In addition the graph will contain 
squares coming from (2,2)-tetrahedra. This class of graphs 
can be described by a two-matrix model:
\beq\label{3.1}
Z \equ \int \!\!\d A \d B \, \e^{ \mi N \tr \left( (A^2\pl B^2) 
\mi \a (A^3 \pl B^3)\mi \b ABAB\right)}.
\eeq
$A$ and $B$ are $N\times N$ Hermitean matrices and the special 
graphs are selected in the large $N$ limit. The coupling 
constants $\a,\b$ can be related to the gravitational coupling 
constants. The phase diagram is shown in Fig.\ \ref{fig4} (for a slightly 
different model, see \cite{ajlv} for details). The matrix model 
is defined for small values of $\a,\b$ and has a critical line.
This line corresponds to the continuum limit of the 3d quantum 
gravity theory. Small values of $\b$ on the critical line correspond 
to small values of the bare gravitational coupling constant.
Increasing $\b$ or the gravitational coupling constant
we meet a phase transition, separating the weak coupling 
phase we have observed in the computer simulations from
a strong coupling phase. The matrix model allows 
spherical graphs at $t \pl a/2$ such that the (1,3)- and (3,1)-triangles no
longer correspond to triangulations of $S^2$, but rather a 
disconnected set of baby universe. One can say that gravity  
becomes so strong that space is torn apart into many pieces 
connected by thin wormholes, \cite{ajlv}. In this way
3d Lorentzian dynamical triangulations provide us with 
an explicit realization of the quantum foam ideas of 
Hawking and Wheeler.

\end{document}